\documentstyle{article}
\title{\LARGE To the question of the integration of Plebansky Equation}
\author{A.~N.~Leznov$\footnote{e-mail: andrey@buzon.uaem.mx}$\thanks{ Universidad Autonoma del Estado de Morelos, CCICAp,Cuernavaca, Mexico}} \date{}

\newcommand{\be}{\beta}

\begin{document}
\maketitle

\begin{abstract}
It is shown that corresponding to Plebansky equation of symmetry posses the infinite set of solutions, which we present in explicit form. This fact leads to conclusion about possibility to find series solutions of the Plebansky equation in analytic form. Some classes of explicit solution are presented. 
\end{abstract}

\section{Introduction}

The goal of the present paper is to investigate Plebansky equation on 
the subject of its integrability. The Plebansky equation is 
non homogeneous complex Monge-Ampher equation
\begin{equation}
Det_n (\phi_{x_i,\bar x_j})=1\label{MA}
\end{equation}
in the case $n=2$. $\phi$ is unknown function of $2n$ independent
variables $x_i,\bar x_j$.

Equation of Plebansky arise in self-dual gravity \cite{1} and we will use it in the form
\begin{equation} 
v_{y,\bar y}v_{z,\bar z}-v_{y,\bar z}v_{z,\bar y}=1\label{PL}
\end{equation}
which will call as a basic form of this equation.

At this place we would like to notice that the general solution of 
homogeneous Monge-Ampher equation (with zero instead of the unity in
the right side of (\ref{MA})), was recently found in implicit form in \cite{MA}. 

We remind to the reader some known facts from the theory of differential equations \cite{O}.
With each system of differential equations it is connected symmetry system of equations, which arise after differentiation of the initial system by some parameter and denotation such obtained derivative $\dot a=A$ as a new unknown function. If linear system of such equations have exact solution on the class of solutions of the initial system the last one is exactly integrable. This means that it is possible to present its solution (in explicit or implicit form) depending on necessary number of arbitrary functions sufficient for the statement the problem of Cauchy-Kovalevskai. Such solution is called as the general one. If there exist series solutions of the symmetry equation then initial system poses the same number of exact solutions, but each solution is not the general one.  

In the present paper we would like to show that in the case of Plebansky equation second possibility take place. It may be found infinite series solutions of its symmetry equation but not the general one. Thus it is not possible to find general solution of this equation in analytic form. But infinite number of its exact solutions exist and the goal of the present paper consists in description the ways for obtaining them.

\section{Preliminary manipulations}

Now we would like to rewrite (\ref{PL}) in equivalent form
\begin{equation} 
(v_{\bar y}v_{z,\bar z}-v_{\bar z}v_{z,\bar y})_y+ 
(v_{y,\bar y}v_{\bar z}-v_{y,\bar z}v_{\bar y})_z=2\label{SPLI}
\end{equation}
\begin{equation} 
(v_y v_{z,\bar z}-v_z v_{\bar z,y})_{\bar y}+ 
(v_{y,\bar y}v_z-v_{\bar y,z}v_y)_{\bar z}=2\label{SPLII}
\end{equation}
The last equations (\ref{SPLI}) and (\ref{SPLII}) can be partially resolved as
$$
v_{\bar y}v_{z,\bar z}-v_{\bar z}v_{z,\bar y}=\theta_z+y,\quad
v_{y,\bar y}v_{\bar z}-v_{y,\bar z}v_{\bar y}=-\theta_y+z
$$
$$
v_y v_{z,\bar z}-v_z v_{\bar z,y}=\theta_{\bar z}+\bar y, ,\quad v_{y,\bar y}v_z-v_{\bar y,z}v_y=
-\theta_{\bar y}+\bar z
$$
It is important to notice that function $\theta$ satisfy the
the symmetry equation corresponding to Plebansky one \cite{O}.

Resolving the last equalities with respect to $v_{\bar y},v_{\bar z}$ (or to $v_y,v_z$), taking into  account the Plebansky equation (\ref{PL}) for $v$ we obtain
\begin{equation}
v_{\bar y}=v_{y,\bar y}\theta_z-v_{z,\bar y}\theta_y+z v_{z\bar y}+
y v_{y,\bar y} ,\quad 
v_{\bar z}=v_{y,\bar z}\theta_z-v_{z,\bar z}\theta_y+z v_{z\bar z}+
y v_{y,\bar z} \label{FS}
\end{equation}
\begin{equation}
v_y=v_{y,\bar y}\theta_{\bar z}-v_{y,\bar z}\theta_{\bar y}+\bar z v_{y,\bar z}+
\bar y v_{y,\bar y} ,\quad 
v_z=v_{z,\bar y}\theta_{\bar z}-v_{z,\bar z}\theta_{\bar y}+\bar z v_{z\bar z}+
\bar y v_{z,\bar y} \label{FSI}
\end{equation}
Equating second mixed derivatives of $v$ with respect to bar variables (or not to the bar ones), we conclude that function $\theta$ in both cases is the solution of the symmetry equation corresponding to (\ref{PL}) namely
\begin{equation}
v_{\bar y,y}\theta_{z,\bar z}+\theta_{\bar y,y}v_{z,\bar z}- 
v_{y,\bar z}\theta_{\bar y, z}-\theta_{y,\bar z}v_{\bar y,z}=0 \label{SE}
\end{equation}

\section{Recurrent formula for solution of the symmetry equation}

Rewriting (\ref{SE}) in two equivalent forms
\begin{equation}
v_{\bar y,y}\theta_z-v_{\bar y,z}\theta_y=\tilde \theta_{\bar y},\quad
v_{\bar z,y}\theta_z-v_{\bar z,z}\theta_y=\tilde \theta_{\bar z}\label{SE1}
\end{equation}
\begin{equation}
v_{\bar y,y}\theta_{\bar z}-v_{\bar z,y}\theta_{\bar y}=\tilde \theta_y,\quad
v_{\bar y,z}\theta_{\bar z}-v_{\bar z,z}\theta_{\bar y}=\tilde \theta_z \label{SE2}
\end{equation}
By the same way as it was done in the previous section it is possible to check that $\tilde \theta$ functions from (\ref{SE1}),(\ref{SE2}) satisfy the symmetry equation (\ref{SE}).
Obvious solution of symmetry equation is derivatives of $v$ with respect to 4 independent arguments of the problem $y,z,\bar y,\bar z$. Thus with help of (\ref{SE1}),(\ref{SE2}) it is possible to construct infinite serie solutions of symmetry equation (compare with \cite{LYM}).

\subsection{Static case}

In this subsection we would like explain the case when symmetry equation has a exact solution and how this connected with the integrable property of the initial system.

Let us consider "time independent" configurations, when $v=v(z+\bar z,y,\bar y)$. 
In this case the Plebanski equation (\ref{PL}) and main symmetry equation  (\ref{FS}) 
are reduced correspondingly
$$
v_{y,\bar y}v_{z,z}-v_{y,z}v_{z,\bar y}=1
$$
$$
v_{\bar y,y}\theta_z-v_{\bar y,z}\theta_y=\tilde \theta_{\bar y},\quad
v_{z,y}\theta_z-v_{z,z}\theta_y=\tilde \theta_z
$$
Let us sick solution of the symmetry equation in terms of three independent 
coordinates $(v_z,y,\bar y)$. The last symmetry system in these coordinates looks as
$$
\theta_{v_z}v_{z,\bar y}+\theta_{\bar y}=\tilde \theta_{v_z}-v_{z,\bar y}\tilde \theta,\quad
\theta_{v_z}v_{z,z}=-v_{z,z}\tilde \theta_y
$$
From which follows that functions $\tilde \theta,\theta$ are connected by the linear system of equations
$$
\tilde \theta_{\bar y}=\theta_{v_z},\quad \tilde \theta_{v_z}=-\theta_y 
$$
which is equivalent to three dimensional Laplace equation. Thus in the static case
the symmetry equation possesses the general solution and thus in this case Plebansky 
equation is exactly integrable. 

Below we present its general solution in implicit form
\begin{equation}
v_{\bar y}=L_{v_z}(v_z,y,\bar y)\quad z=L_y(v_z,y,\bar y) \label{SPL}
\end{equation}
where function $L$ satisfy Laplace equation in three dimension
$L_{v_z,v_z}+L_{y,\bar y}=0$. General solution of Laplace equation in three dimension depend on two arbitrary functions of two arguments. And thus constructed above is a general solution of Plebansky equation in static case.

This solution may be obtained directly. Indeed Plebansky equation in this case may be rewritten as
$$
(v_{\bar y})_y (v_z)_z-(v_z)_y (v_{\bar y})_z=1
$$
and functions $(v_{\bar y},v_z)$ may be considered as transformed impulse and coordinate with unity Poisson bracket between them.
This system is resolved by canonical transformation with generating function $G(v_z,y,\bar y)$
by usual formulae
$$  
v_{\bar y}=G_{v_z}(v_z,y,\bar y),\quad z=G_y(v_z,y,\bar y)
$$
From condition of equality of the second mixed derivatives of $v$ function we conclude that $G$
satisfy Laplace equation in three dimension. This is exactly  (\ref{SPL}) above.

\section{General strategy and example explaining it}

General (\ref{FS}) is the system of two equations on two unknown functions $v,\theta$. As a consequence we know that $v$ function satisfy Plebansky equation and $\theta$ function the symmetry one. But all solutions of symmetry equation are enumerated in the previous section.
And thus if we change $\theta$ in (\ref{FS}) on one of solution of the previous section we will
obtain the self consistent system of two equations only on one unknown function $v$.
Solution of this system (if it will be possible to find it) will be solution of Plebansky equation corresponding to such chose solution of symmetry equation.

To show that such idea is not mean less we at first consider simple example of (\ref{FS}) 
under the chose $\theta=v_{\bar y}$. But result of solution of Plebansky equation will be absolutely non trivial one. 

We rewrite (\ref{FS}) 
$$
v_{\bar y}=(z \frac{\partial}{\partial z}+y\frac{\partial}{\partial y})v_{\bar y} ,\quad 
v_{\bar z}=1+(z \frac{\partial}{\partial z}+y\frac{\partial}{\partial y})v_{\bar z}
$$
Two independent ordinary differential equations have obvious solution
$$
v_{\bar y}=(y z)^{1\over 2}X_{\bar y}(d,\bar z,\bar y),\quad
v_{\bar z}=1+(y z)^{1\over 2}X_{\bar z}(d,\bar z,\bar y), \quad d\equiv {z\over y}
$$
Further
$$
v_{\bar y,y}={1\over 2} d^{1\over 2}X_{\bar y}-d^{3\over 2}X_{\bar y,d},\quad
v_{\bar y,z}={1\over 2} d^{-1\over 2}X_{\bar y}+d^{1\over 2}X_{\bar y,d}
$$
$$
v_{\bar z,y}={1\over 2} d^{1\over 2}X_{\bar z}-d^{3\over 2}X_{\bar z,d},\quad
v_{\bar z,z}={1\over 2} d^{-1\over 2}X_{\bar z}+d^{1\over 2}X_{\bar z,d}
$$
And equation of Plebansky (\ref{PL}) takes the form
\begin{equation}
X_{\bar y}X_{\bar z,D}-X_{\bar z}X_{\bar y,D}=1,\quad D\equiv \ln d \label{ME}
\end{equation}
Resolving of the last equation is connected with second order ordinary differential equations of the form $X_{D,D}=F(X,D)$ ($F$ arbitrary functions of its arguments). Indeed solution of this equation depends on two arbitrary parameters $c_1,c_2$ and may be represented as some function depending on 3 arguments $X=X(D,c_1,c_2)$.
Let us differential equation for $X$ function argumentson parameters $c_i$. We have $X_{D,D,c_i}=F_X(X,D)X_{c_i}$. From the last equality we conclude $(X_{D,c_1}X_{c_2}-X_{D,c_2}X_{c_1})_D=0$. We will assume that this value is equal to unity (this is always possible to do by corresponding canonical transformation). Now we will consider $c_1,c_2$ as arbitrary functions of the arguments $\bar y,\bar z$. Then equation (\ref{ME}) Looks as
$$        
(X_{c_1}X_{c_2,D}-X_{c_2}X_{c_1,D})((c_1)_{\bar y}(c_2)_{\bar z}-(c_1)_{\bar z}(c_2)_{\bar y})=
((c_1)_{\bar y}(c_2)_{\bar z}-(c_1)_{\bar z}(c_2)_{\bar y})=1
$$
Thus obtained solution of initial Plebansky equation depend on two arbitrary function $F(X,D)$
defined arbitrary equation of the second order and generating function of canonical transformation
resolving the last equation for $c_1,c_2$ functions.

We present the second way for solution equation (\ref{ME}). Let us consider $X_D,X$ as canonical transformed coordinate and impulse variable and $\bar z, \bar y$ as the same initial ones. Then generating function of canonical transformation $W=W(X,\bar y,D)$ satisfy the equations
$$
X_D=W_X(X,\bar y,D) ,\quad   \bar z=W_{\bar y}(X,\bar y,D)
$$
From the second equation we have $X_D=-{W_{\bar y,D}\over W_{\bar y,X}}$ and after substitution into the first equation we obtain
$$
(W_D+{W_X^2\over 2})_{\bar y}=0,\quad W_D+{W_X^2\over 2}=F(X,D)
$$
The second one is Hamilton-Jacobi equation of the particle motion in potential field $V=F(X,D)$, It leads to second order differential equation considered above. Hamilton-Jacobi equation reduce the number of independent variables in generating function $W$ from 3 up to 2. And thus solution of Plebansky equation is determined by two functions $W,F$ each one of two independent variables.

\section{Equation of Plebansky in non usual variables}

Let us introduce notations $R=\ln v_{\bar y}+\ln v_{\bar z},\Delta=
\ln v_{\bar y}-\ln v_{\bar z}$. Or in other words $v_{\bar y}=\exp ({R+\Delta\over 2}),
v_{\bar z}=\exp ({R-\Delta\over 2})$. In these notations the pair of equations (\ref{LS})
take the form
\begin{equation}
R_y,\theta_z-R_z\theta_y+z R_z+y R_y=2 ,\quad 
\Delta_y\theta_z-\Delta_z\theta_y+z\Delta_z+y\Delta_y=0 \label{FSI}
\end{equation}
In the last equations let us pass from independent variables $y,z$ to independent variables
$\theta,d={z\over y}$. Corresponding necessary formulae are presented below
\begin{equation}
y=Y(\theta,d,\bar y,\bar z),\quad 1=Y_{\theta} \theta_y-Y_d{z\over y^2},\quad 0=Y_{\theta} \theta_z+
Y_d{1\over y}\label{TF}
\end{equation}
$$ 
R_y==R_{\theta} \theta_y-R_d{z\over y^2},\quad R_z=R_{\theta} \theta_z+
R_d{1\over y}
$$
and the same formulae for derivatives of the $\Delta$ function. In all relations above its 
necessary to keep in mind that all function under consideration depend also on bar arguments
$\bar y,\bar z$. Such dependence will be taken into account on some forward steps of
calculations.

In variables $\theta,d$ the system (\ref{FSI}) looks as
\begin{equation}
(Y^2)_{\theta}-Y^2 R_{\theta}=-R_d,\quad (Y^2e^{-R})_{\theta}=(e^{-R})_d,\quad
Y^2={\Delta_d\over \Delta_{\theta}}\label{MS}
\end{equation}

To do the last equality more symmetrical to $y,z$ variables let us multiply the last one
on $d={z\over y}$. We have in a consequence
\begin{equation}
d Y^2=y z={d\Delta_d\over \Delta_{\theta}}={\Delta_D\over \Delta_{\theta}},\quad D=\ln d
\label{?}
\end{equation}
Now we would like to satisfy equation of Plebansky
$$
v_{y,\bar y}=({R_y+\Delta_y\over 2})\exp ({R+\Delta\over 2}),\quad
v_{z,\bar y}=({R_z+\Delta_z\over 2})\exp ({R+\Delta\over 2})
$$
$$
v_{y,\bar z}=({R_y-\Delta_y\over 2})\exp ({R-\Delta\over 2}),\quad
v_{z,\bar z}=({R_z-\Delta_z\over 2})\exp ({R-\Delta\over 2})
$$
And thus the equation of Plebansky looks as
$$
e^R(R_z\Delta_y-R_y\Delta_z)=2 
$$
In the last equation let us pass to variables $\theta,d$ with the help of above 
formulae. We obtain
$$
R_d\Delta_{\theta}-R_{\theta}\Delta_d=e^{-R}(Y^2)_{\theta}
$$
Let us compare the last equation with obtained above (\ref{MS}) ones. As a direct corollary
we have
\begin{equation}
e^{-R}=-\Delta_{\theta}\label{!!}
\end{equation}
The last relations have as a direct consequence two first relations (\ref{MS}).

In all calculations above no information about dependence of all functions involved 
with respect to bar arguments was not used. Now let use condition of equality of the second 
mixed derivatives of $v$ function with respect to $\bar y,\bar z$ arguments in 
notations introduced above.
$$
(\exp ({R+\Delta\over 2}))_{\bar z}=(\exp ({R-\Delta\over 2}))_{\bar y},\quad
\exp \Delta (R_{\bar z}+\Delta_{\bar z})=(R_{\bar y}-\Delta_{\bar y})
$$
But functions $R,\Delta$ are connected by (\ref{!!}) and function $\Delta$ depends in its turn on 
$\theta$ and thus terms with derivatives on bar arguments look as (we present left hand side term)
\begin{equation}
-(\ln \Delta_{\theta})_{\bar z}+\Delta_{\bar z}+(-(\ln \Delta_{\theta})_{\theta}+\Delta_{\theta})
\theta_{\bar z} \label{??}
\end{equation}
After some computations using (\ref{?}) we come to equation,which function $\Delta$ satisfy
\begin{equation}
e^{5\Delta}((e^{-\Delta})_{\theta,\bar z}(e^{-\Delta})_{\theta,d}-
(e^{-\Delta})_{\theta,\theta}(e^{-\Delta})_{\bar z ,d})=
(e^{\Delta})_{\theta,\bar y}(e^{\Delta})_{\theta,d}-
(e^{\Delta})_{\theta,\theta}(e^{\Delta})_{\bar y,d} \label{GR1}
\end{equation}
In the case when symmetry function depends only from non bar variables equation (\ref{GR1}) looks much more simple (in (\ref{??}) $\theta_{\bar z}=\theta_{\bar y}=0$)
$$
e^{3\Delta}(e^{-\Delta})_{\theta,\bar z}=(e^{\Delta})_{\theta,\bar y}
$$
This equation is equivalent to equation considered in section 4. It will be explained in one of sections below.

After introduction new function $\theta=\Theta (\Delta, d,\bar y,\bar z)$ it looks as
\begin{equation}
e^{\Delta}(\Theta_{\Delta,\Delta}\Theta_{d,\bar z}-\Theta_{\Delta,d}\Theta_{\Delta,\bar z}+
\Theta_{\Delta}\Theta_{d,\bar z})=
\Theta_{\Delta,\Delta}\Theta_{d,\bar y}-\Theta_{\Delta,d}\Theta_{\Delta,\bar y})-
\Theta_{\Delta}\Theta_{d,\bar y} \label{GR2}
\end{equation}
or in variables $d=D(\Delta,\theta,\bar y, \bar z)$ it looks as
\begin{equation}
e^{\Delta}Det_3 \pmatrix{ D_{\bar z} & D_{\Delta} & D_{\theta} \cr
D_{\Delta, \bar z} & D_{\Delta,\Delta}+D_{\Delta} & D_{\Delta\theta} \cr
D_{\theta, \bar z} & D_{\theta,\Delta} & D_{\theta,\theta} \cr}=
Det_3 \pmatrix{ D_{\bar y} & D_{\Delta} & D_{\theta} \cr
D_{\Delta, \bar y} & D_{\Delta,\Delta}-D_{\Delta} & D_{\Delta\theta} \cr
D_{\theta, \bar y} & D_{\theta,\Delta} & D_{\theta,\theta} \cr} \label{GR3}
\end{equation}
Each of three equations above are equivalent to the initial Plebansky equation (\ref{PL}).

\subsection{Equation (\ref{GR2}) in the integral motion form}

Let us introduce the following notation the operators of the differentiations
$$
L^{\pm}=e^{{\Delta\over 2}}{\frac {\partial}{\partial \bar z}} \pm e^{-{\Delta\over 2}}{\frac 
{\partial}{\partial \bar y}},\quad L^0={\frac {\partial}{\partial \Delta}}
$$
with the obvious commutation relations
$$
[L^0,L^{\pm}]={1\over 2} L^{\mp},\quad [L^+,L^-]=0
$$
(these are commutation relation of algebra of two dimension plane).
In these notations equation (\ref{GR2}) may be rewritten as
$$
\Theta_{\Delta,\Delta}(L^-\Theta)_d+2\Theta_{\Delta}(L^-\Theta_d)_{\Delta}-
2\Theta_{\Delta}(L^-\Theta_{\Delta})_d-\Theta_{\Delta,d}(L^-\Theta_{\Delta})=0
$$
Or after dividing on $(\Theta_{\Delta})^{{1\over 2}}$ and trivial regrouping of the terms we
obtain
\begin{equation}
((\Theta_{\Delta})^{{1\over 2}}(L^-\Theta_d))_{\Delta}=((\Theta_{\Delta})^{{1\over 2}}
(L^-\Theta_{\Delta}))_d \label{GR2A}
\end{equation}
Or (\ref{GR2}) may be rewritten in integral of motion form. We remind the reader that all equations (\ref{GR1}),(\ref{GR2}), (\ref{GR3}) where obtained from the equality of the 
derivatives on the bar variables.

\section{Some examples of particular solutions}

In this section we present some particular solutions of equations (\ref{GR2}) or equivalent to it
(\ref{GR1}), (\ref{GR3}). From this consideration it will be clear that these equation leads
indeed to solution of initial Plebansky equation.

\subsection{The case when symmetry function do not depend on bar variables}

From explicit form of symmetry equation of the second section it is clear that simplest obvious  its solution is $\theta=\theta (y,z)$. Equation describing this situation are the following ones
$$
e^{3\Delta}(e^{-\Delta})_{\theta,\bar z}=(e^{\Delta})_{\theta,\bar y},\quad {\Delta_d\over \Delta_{\theta}}=Y^2(\theta,D)=y z
$$
From the second equation that function $e^{\Delta}$ really is the function of only three variables
$\bar y,\bar z, g(\theta,D)$. Let us seek solution of the first equation in a form
$$
e^{\Delta}=-{X_{\bar y}\over X_{\bar z}}
$$
where $X=X(\bar y,\bar z, g(\theta,D))$. 
After such substitution first equation takes the form 
$$
({X_{\bar y}\over X_{\bar z}})^3({X_{\bar y,g}X_{\bar z}-X_{\bar z,g}X_{\bar y}\over X^2_{\bar y}})_{\bar z}=({X_{\bar y,g}X_{\bar z}-X_{\bar z,g}X_{\bar y}\over X^2_{\bar z}})_{\bar y}
$$
From the last equality it follows immediately
$$
X_{\bar y,g}X_{\bar z}-X_{\bar z,g}X_{\bar y}=F(X,g) \to 1
$$
where $F$ arbitrary function of its two arguments. At last by substitution $X\to Y(X,g)$ it is possible to equate $F$ to unity and we come back to equation considered in 4 section.
Solution of Plebansky equation id the following one
$$
v_{\bar y}=\exp ({R+\Delta\over 2}=({\exp \Delta\over -\Delta_{\theta}})^{{1\over 2}}={g_{\theta}}^{-{1\over 2}}X_{\bar y}, \quad v_{\bar z}={g_{\theta}}^{-{1\over 2}}X_{\bar z},\quad
v={g_{\theta}}^{-{1\over 2}}X
$$

\subsection{Solution do not depend on one bar coordinate}

Let us assume that solution of equation (\ref{GR2}) does not depend on $\bar z$ coordinate. Then equation for $\Theta$ function takes the form
$$
e^{-\Delta}(e^{-\Delta}\Theta_{\Delta})_{\Delta}\Theta_{D,\bar y}-e^{-\Delta}\Theta_{\Delta,D}e^{-\Delta}\Theta_{\Delta,\bar y})=0
$$
This is exactly three dimensional subclass of complex Monge-Ampher (four dimensional equation),
solution of which in implicit form is known \cite{MA}. 

It may be expressed via (in terms of) $\psi(\Delta,d,\bar y)$ function which is solution of the equation 
\begin{equation}
e^{\Delta}+F_{\psi}(D,\psi)+\bar F_{\psi}(\bar y,\psi)=0 \label{BUIT}
\end{equation}
In the last equation for $\psi(\Delta,D,\bar y)$ 
two arbitrary functions  $F(D,\psi),\bar F(\bar y,\psi)$ assumed to be known. Of course  solution of this equation (in general case) may be obtained only in implicit form.

After this solution of the equation for $\Theta (D,\bar y, \Delta)$ function is resolves as follows
$$
\Theta_{\Delta}=e^{\Delta} \psi,\quad \Delta_{\Theta}=e^{-\Delta} \psi^{-1},\quad \Theta_{\bar y}=-\bar F_{\bar y},\quad \Theta_D=F_D
$$
And this is a general solution of this equation.
 
The solution of the initial equation of Plebansky in coordinates $z,y,\bar z,\bar y$ is given
by equations
\begin{equation}
v_{\bar y}=e^{\Delta} (\psi)^{{1\over 2}},\quad v_{\bar z}=(\psi)^{{1\over 2}},\quad
yz+F_D(D,\psi)=0 \label{MA3}
\end{equation}
We present below direct checking of the formulae above. Equation of Plebansky in its basic form after substitution (\ref{MA3})  looks as 
$$
{1\over 2}(e^{\Delta}_y\psi_z-e^{\Delta}_z\psi_y)=-{1\over 2}F_{\psi,D}{y\psi_y+z\psi_z\over y z}=1
$$
In the last step of calculation necessary use last formulae of (\ref{MA3})  
$$
y+F_{D D}{1\over z}+F_{D,\psi} \psi_z=0,\quad z-F_{D D}{1\over y}+F_{D,\psi} \psi_y=0,\quad
D=\ln z-\ln y
$$

\subsubsection{Solution of the example from a section 4}

Let us sick solution of (\ref{GR2}), (\ref{GR2A}) in a form (of course such form is a direct consequence of calculations of the 4 section)
$$
\Theta=(e^{-\Delta}-r(D,\bar y,\bar z))^{-1}
$$
All necessary derivatives are the following ones
$$
\Theta_{\Delta}=e^{-{1\over 2}\Delta}\Theta^2,\quad \Theta_D=r_D\Theta^2,\quad \Theta_{\bar y}=r_{\bar y}\Theta^2,\quad \Theta_{\bar z}=r_{\bar z}\Theta^2
$$
After calculation of second order derivatives and substitution into (\ref{GR2A}) we obtain equation for $r$ function 
$$
r^3({1\over r})_{D,\bar y}=r_{D,\bar z}
$$
This equation is exactly integrable (author have not met it in literature before) with solution
$$
r=-{X_{\bar z}\over X_{\bar y}}
$$ 
where $X(D,\bar y,\bar z)$ is exactly the function from the section 4. 
Indeed $r_D=-{X_{D,\bar z}X_{\bar y}-X_{D,\bar y}X_{\bar z}\over X^2_{\bar y}}={1\over X^2_{\bar y}}$. 

\subsubsection{Once more possible particular solution}

Let us sick solution of (\ref{GR2}), (\ref{GR2A}) in a form
$$
\Theta=P(\bar y,\bar z, \Delta)+Q(d,\Delta)
$$
After substitution into corresponding equations we come to the linear equation for determining
of the $P_{\Delta}$ function
$$
(e^{-{\Delta\over 2}}{\frac {\partial}{\partial \bar y}}-e^{{\Delta\over 2}}{\frac {\partial}
{\partial \bar z}}) P_{\Delta}=0
$$
with the obvious solution
$$
P=\int^{\Delta} d\delta p(e^{{-\delta\over 2}}\bar z+e^{{\delta\over 2}}\bar y,\delta)
$$
But really for solution of Plebansky equation it is necessary only derivative of $P$ function with respect to the $\Delta$ argument. Indeed
$$
v_{\bar y}=({e^{\Delta}\over \Delta_{\theta}})^{{1\over 2}},\quad v_{\bar z }=({e^{-\Delta}\over \Delta_{\theta}})^{{1\over 2}},\quad \Delta_{\theta}={1\over \Theta_{\Delta}}={1\over P_{\Delta}+Q_{\Delta}}  
$$
and at last connection between $\theta$ and $\Delta$ is given by relation $yz={\Delta_D\over \Delta_{\theta}}=-\Theta_D=-Q_D$.

\section{Discrete transformation}

Let us rewrite main equations of preliminary section once more
\begin{equation}
v_{\bar y}=v_{y,\bar y}\theta_z-v_{z,\bar y}\theta_y+z v_{z\bar y}+
y v_{y,\bar y} ,\quad 
v_{\bar z}=v_{y,\bar z}\theta_z-v_{z,\bar z}\theta_y+z v_{z\bar z}+
y v_{y,\bar z} \label{FSA}
\end{equation}
and pay attention that in connection with (\ref{SE1}) these equations may be rewritten as
$$
v_{\bar y}=\tilde \theta_{\bar y}+z v_{z,\bar y}+y v_{y,\bar y},\quad v_{\bar z}=\tilde \theta_{\bar z}+z v_{z,\bar z}+y v_{y,\bar z}
$$
The last equation may resolved with respect $\tilde \theta$ function
$\tilde \theta=v-z v_z-y v_y$. And thus after substitution this expression for $\tilde v$ function we obtain linear system of equations for $\tilde v$ in a form 
\begin{equation}
\tilde v_{\bar y}=(-(O v_z) \frac {\partial}{\partial y}+(O v_y) \frac {\partial}{\partial z}+ O) \tilde v_ {\bar y} ,\quad \tilde v_{\bar y}=(-(O v_z) \frac {\partial}{\partial y}+(O v_y) \frac {\partial}{\partial z}+ O) \tilde v_ {\bar z} \label{FSAI}
\end{equation}
where operator $O \equiv (y\frac {\partial}{\partial y}+z\frac {\partial}{\partial z})$.
Equations (\ref{FSAI}) is linear system of equations for determining $\tilde v_{\bar y},\tilde v_{\bar z}$ functions by known $v$ solution of Plebansky equation.

At this moment it is unknown to the author the systematic way for resolving (\ref{FSAI}).

\section{Outlook}

The main result of the present paper is a new approach to the problem of Plebansky equation.
This approach allows rewrite Plebansky equation in coordinates involving symmetry function
and find series solutions of this equation. Solutions obtained in such way depends on two arbitrary functions each of two variables. Thus this is not a general solution of Plebansky equation which must depend on two arbitrary function each of three variables. But investigation of symmetry equation show that general solution of Plebansky equation it is not possible present in analytic form.   Situation exactly the same as for instance in the case of famous nonlinear one dimensional Schredinger equation, where it is possible to find infinite series solution of soliton like type but not general one.

Constructed in the present paper solutions are connected with ordinary differential equation
of the second order $ X_{z,z}=F(X,z)$ and it arise very interesting problem to understand
what relation has this equation to group of symmetry of Plebansky equation responsible for its
integrable properties.

\section{Appendix}

In this Appendix we consider simplest example from section $(6.0.1)$ chose arbitrary functions
in such simple form that all calculation are possible to do in explicit form. Let us chose
$F={(d+\psi)^2\over 2},\bar F={(\bar y+\psi)^2\over 2}$. The main equation allow determine
$\psi$ and all other necessary variables in explicit form in usual for Plebansky equation variables
$$
\psi=-{e^{\Delta}+\bar y+d\over 2}, \quad y z={\Delta_d\over\Delta_{\theta}}=-\Theta_d=
-(d+\psi)={e^{\Delta}+\bar y-d\over 2},
$$
$$
 e^{\Delta}=2y z+D-\bar y,\quad \psi=-(yz+d),\quad D=\ln {z\over y}
$$
In connection with results of section $(6.0.1)$ solution of basic Plebansky equation looks as
$$
v=[(2y z+\ln {z\over y})\bar y-{(\bar y)^2\over 2}+\bar z](y z+\ln {z\over y})^{1\over 2}
$$

\end{document}